\documentclass[a4paper,nobibnotes,nofootinbib,superscriptaddress]{revtex4}

\usepackage{amssymb}
\usepackage{amsmath}
\usepackage{epsfig}
\def \et {E_{T}}
\newcommand{\met}{\mbox{$\not\!\!\et$}}
\newcommand{\aZerow}{a_0^W}

\newcommand{\aCw}{a_C^W}
\newcommand{\aCz}{a_C^Z}

\newcommand{\be}{\begin{equation}}
\newcommand{\ee}{\end{equation}}
\newcommand{\bea}{\begin{eqnarray}}
\newcommand{\eea}{\end{eqnarray}}

\newcommand{\g}{\gamma}

\newcommand{\intc}[1]{{\int\frac{d#1}{2i\pi}}}
\newcommand\lr[1]{{\left({#1}\right)}}

\begin{document}

\title{Mueller Navelet jets, jet gap jets and anomalous $WW\gamma \gamma$
couplings in $\gamma$-induced processes at the LHC}
\author{C. Royon}\email{christophe.royon@cea.fr}
\affiliation{IRFU/Service de physique des particules, CEA/Saclay, 91191 Gif-sur-Yvette cedex, France}

\begin{abstract}

We describe two different important measurements to be performed at the LHC.
The Mueller Navelet jet and jet gap jet cross section represent a test of BFKL
dynamics and we perform a NLL calculation of these processes and compare it
with recent Tevatron measurements. The study of the $WW\gamma \gamma$ couplings
at the LHC using the forward detectors proposed in the ATLAS Forward Physics
project as an example allows to probe higgsless and extradimension
models via anomalous quartic
couplings since the reach is improved by four orders of magnitude with respect 
to the LEP results.
\end{abstract}

\maketitle

\section{Mueller Navelet jets at the LHC}
In this section, we give the BFKL NLL cross section calculation for Mueller
Navelet processes at the Tevatron and the LHC. Since the starting point of this
study was the description of forward jet production at HERA, we start by
describing briefly these processes.

\subsection{Forward jets at HERA}

\begin{figure}
\centerline{\includegraphics[width=0.45\columnwidth]{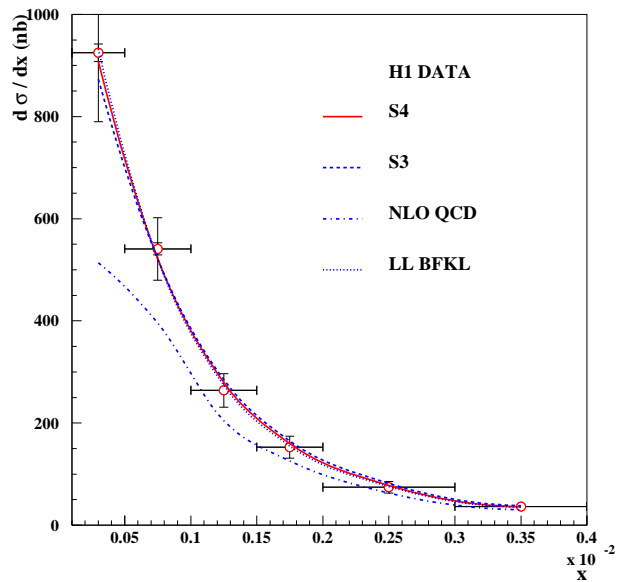}}
\caption{Comparison between the H1 $d \sigma /dx$ measurement 
with predictions for BFKL-LL, BFKL-NLL (S3 and S4 schemes) and DGLAP NLO calculations
(see text). S4, S3 and LL BFKL cannot be distinguished on that figure.}\label{Fig1a}
\end{figure}

Following the successful BFKL \cite{bfkllo} parametrisation of the forward-jet 
cross-section $d\sigma/dx$ at Leading Order (LO)
at HERA \cite{mr,mrb}, it is possible to perform a similar study using Next-to-leading (NLL)
resummed BFKL kernels.
Forward jets at HERA are an ideal observable to look for BFKL resummation
effects. The interval in rapidity between the scattered lepton and the jet in
the forward region is large, and when the photon virtuality $Q^2$ is close to
the transverse jet momentum $k_T$, the DDLAP cross section is small because of
the $k_T$ ordering of the emitted gluons.
In this short report, we will only discuss the
phenomelogical aspects and all detailed calculations can be found in 
Ref.~\cite{fwdjet} for forward jets at HERA and in Ref.~\cite{mnjet} for Mueller Navelet jets
at the Tevatron and the LHC.

\subsection{BFKL NLL formalism}
The BFKL NLL \cite{bfkl} longitudinal transverse cross section reads:
\begin{eqnarray}
\frac{d\sigma^{\gamma*p\!\rightarrow\!JX}_{T,L}}{dx_Jdk_T^2}=
\frac{\alpha_s(k_T^2)\alpha_s(Q^2)}{k_T^2Q^2}\ f_{eff}(x_J,k_T^2)
\int d\gamma \left(\frac{Q^2}{k_T^2}\right)^{\gamma} \phi^{\gamma}_{T,L}(\gamma)\ 
e^{\bar\alpha(k_T Q)\chi_{eff}[\gamma,\bar\alpha(k_T Q)]Y}
\label{nll}
\end{eqnarray}
where $x_J$ is the proton momentum fraction carried by the forward jet,
$\chi_{eff}$ is the effective BFKL NLL kernel and the $\phi$s are 
the transverse and longitunal impact factors taken at LL. The effective kernel
$\chi_{eff}(\gamma,\bar\alpha)$ is
defined from the NLL kernel $\chi_{NLL}(\gamma,\omega)$ by solving the implicit 
equation numerically
\begin{eqnarray}
\chi_{eff}(\gamma,\bar\alpha)=\chi_{NLL}\left[\gamma,\bar\alpha\ 
\chi_{eff}(\gamma,\bar\alpha)\right]\ ,
\label{eff}
\end{eqnarray}

The integration over $\gamma$ in Eq.~\ref{nll} is performed numerically.
It is possible to fit directly $d \sigma/dx$ measured by the H1 collaboration
using this formalism with one single parameter, the normalisation.
The values of $\chi_{NLL}$ are
taken at NLL~\cite{bfkl} using different resummation schemes to remove spurious
singularities defined as S3 and S4~\cite{resum}. Contrary to LL BFKL, it is
worth noticing that the coupling constant $\alpha_S$ is taken using the
renormalisation group equations, the only free parameter in the fit being the
normalisation.

\begin{figure}
%\begin{wrapfigure}{r}{0.5\columnwidth}
\centerline{\includegraphics[width=0.75\columnwidth]{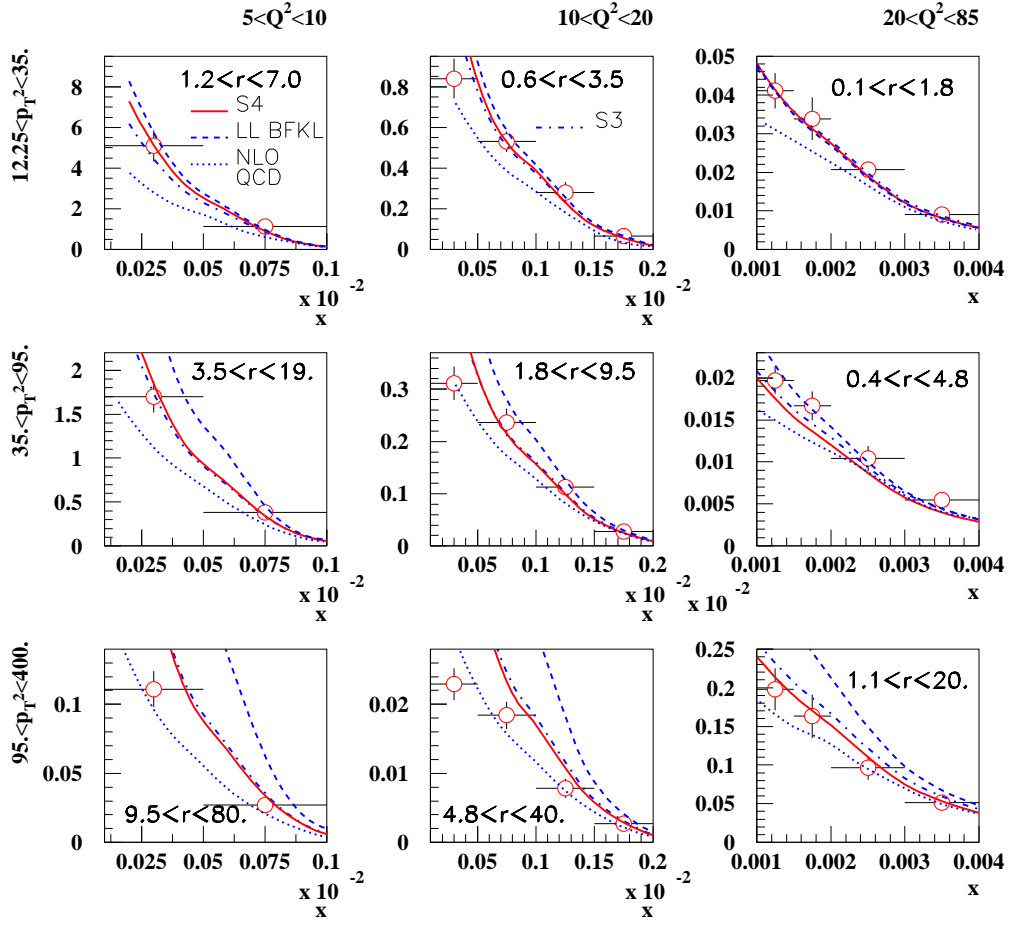}}
\caption{Comparison between the H1 measurement of the triple differential cross
section with predictions for BFKL-LL, BFKL-NLL and DGLAP NLO calculations
(see text).}\label{Fig2}
%\end{wrapfigure}
\end{figure}

To compute $d \sigma/dx$ in the experimental bins, we need to integrate
the differential cross section on the bin size in $Q^2$, $x_{J}$ (the momentum
fraction of the proton carried by the forward jet), $k_T$ ,
while taking into account the experimental cuts. To simplify the numerical
calculation, we perform the integration on the bin using the
variables where the cross section does not change rapidly, namely $k_T^2/Q^2$,
$\log 1/x_{J}$, and $1/Q^2$. Experimental cuts are treated directly at the
integral level (the cut on $0.5<k_T^2/Q^2<5$ for instance) or using a toy Monte
Carlo. 
More detail can be found about the fitting procedure 
in Appendix A of Ref.~\cite{mrb}.

The NLL fits~\cite{fwdjet} can nicely
describe the H1 data~\cite{h1} for the S4  and S3 schemes~\cite{mr,mrb,fwdjet} ($\chi^2=0.48/5$
and  $\chi^2=1.15/5$ respectively
per degree
of freedom with statistical and systematic errors added in quadrature).
The curve using a LL fit is indistinguishable in Fig.~\ref{Fig1a} from the result of the
BFKL-NLL fit.
The DGLAP NLO calculation fails to
describe the H1 data at lowest $x$ (see Fig.~\ref{Fig1a}). We also checked the effect
of changing the scale in the exponential of Eq.~\ref{nll} from $k_TQ$ to
$2k_TQ$ or $k_TQ/2$ which leads to a difference of 20\% on the cross section while
changing the scale to $k_T^2$ or $Q^2$ modifies the result by less than 5\%
which is due to the cut on $0.5 < K_T^2/Q^2<5$. Implementing the higher-order
corrections in the impact factor due to exact gluon dynamics in the $\gamma^*
\rightarrow q \bar{q}$ transition~\cite{robi} changes the result by less than
3\%.

The H1 collaboration also measured the forward jet triple differential cross
section~\cite{h1} and the results are given
in Fig.~\ref{Fig2}. We keep the same normalisation coming from the fit to $d \sigma/dx$
to predict the triple differential cross section.
The BFKL LL formalism leads to a good description 
of the data when $r=k_T^2/Q^2$
is close to 1 and deviates from the data when $r$ is further away from 1. This
effect is expected since DGLAP radiation effects are supposed to occur when
the ratio between
the jet $k_T$ and the virtual photon $Q^2$ are further away from 1. The BFKL 
NLL calculation
including the $Q^2$ evolution via the renormalisation group equation leads to a
good description of the H1 data on the full range. We note that the higher order
corrections are small when $r \sim 1$, when the BFKL effects are
supposed to dominate. By contrast, they are significant as expected when $r$ is different from
one, ie when DGLAP evolution becomes relevant. We notice that the DGLAP NLO calculation
fails to describe the data when $r \sim 1$, or in the region where
BFKL resummation effects are expected to appear. 

In addition, we checked the dependence of our results on the scale taken in the
exponential of Eq.~\ref{nll}. The effect is a change of the cross section of
about 20\% at low $p_T$ increasing to 70\% at highest $p_T$. Taking the correct
gluon kinematics in the impact factor lead as expected to a better description
of the data at high $p_T$~\cite{fwdjet}. 

\subsection{Mueller Navelet jets at the Tevatron and the LHC}

\begin{figure}
\centerline{\includegraphics[width=0.45\columnwidth]{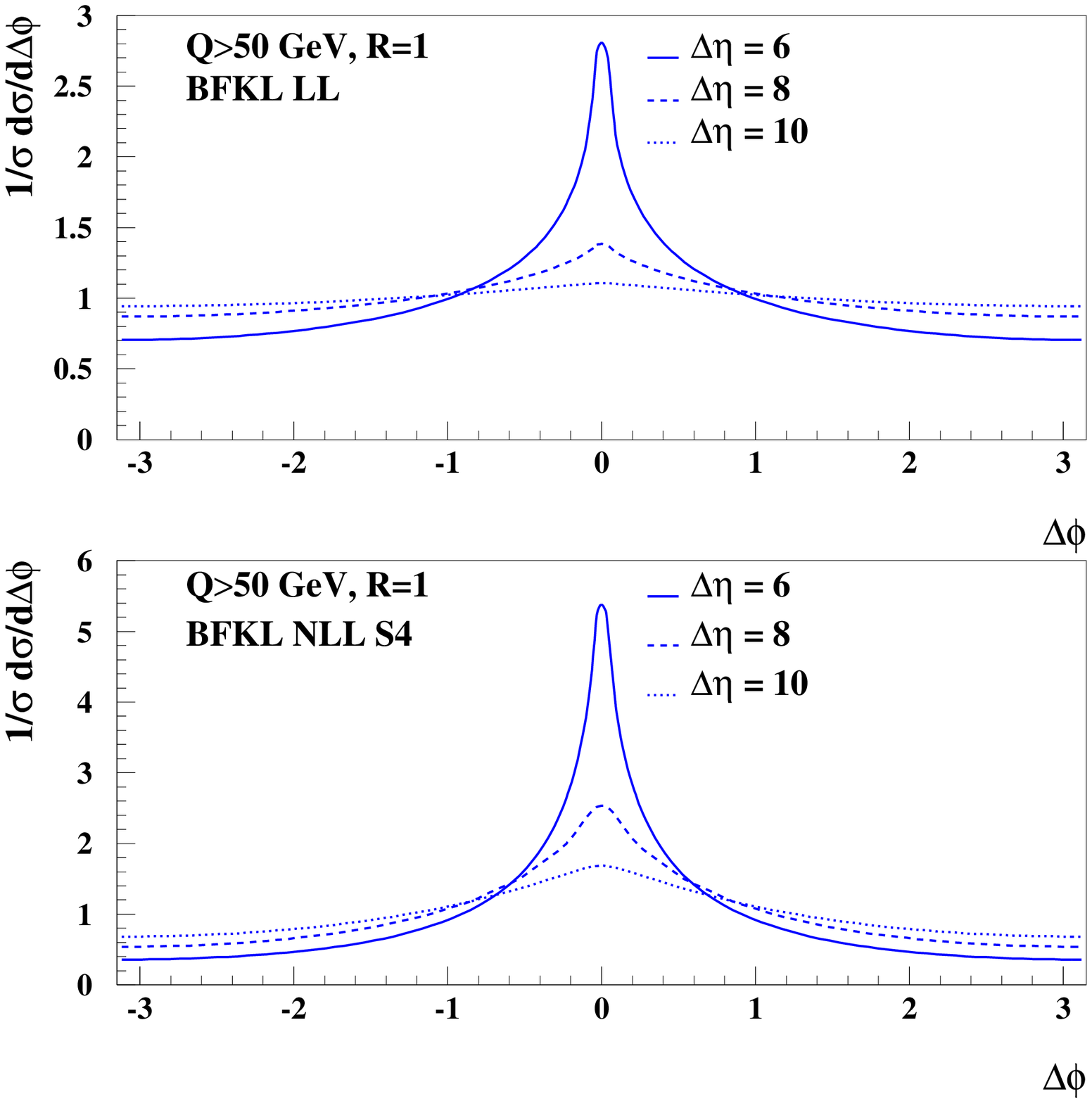}}
%\centerline{\includegraphics[width=0.45\columnwidth]{Fig4b.eps}}
\caption{The Mueller-Navelet jet $\Delta\Phi$ distribution for LHC kinematics in the BFKL framework at 
LL (upper plots) and NLL-S4 (lower plots) accuracy for $\Delta\eta=6,\ 8,\ 10.$}\label{Figlhc}
\end{figure}

\begin{figure}
\centerline{\includegraphics[width=0.45\columnwidth]{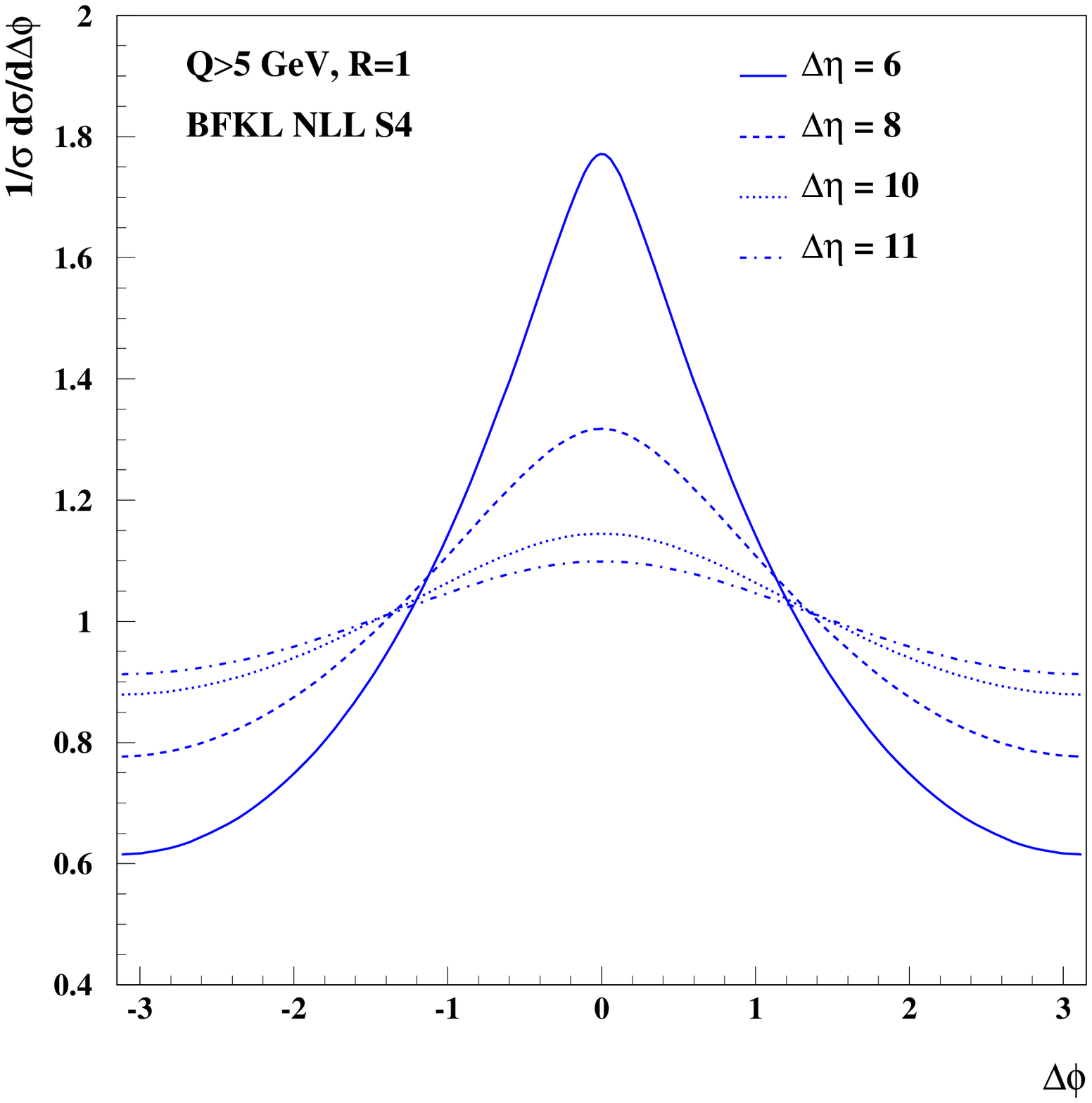}}
\caption{Azimuthal correlations between jets with $\Delta \eta=$6, 8, 10
and 11 and $p_T>5$ GeV in the CDF acceptance. This measurement will represent a
clear test of the BFKL regime.}\label{Fig3}
\end{figure}

Mueller Navelet jets are ideal processes to study BFKL resummation effects~\cite{mn}.
Two jets with a large interval in rapidity and with similar
tranverse momenta are considered. A typical observable to look for BFKL effects
is the measurement of the azimuthal correlations between both jets. The DGLAP
prediction is that this distribution should peak towards $\pi$ - ie jets
are back-to-back- whereas
multi-gluon emission via the BFKL mechanism leads to a smoother distribution.
The relevant variables to look for azimuthal correlations are the following:
\begin{eqnarray}
\Delta \eta &=& y_1 - y_2  \nonumber \\
y &=& (y_1 + y_2)/2 \nonumber \\
Q &=& \sqrt{k_1 k_2} \nonumber \\
R &=& k_2/k_1  \nonumber 
\end{eqnarray}
where $y_{1,2}$ and $k_{1,2}$ are respectively the jet rapidities and transverse
momenta.
The azimuthal correlation for BFKL reads:
\begin{eqnarray}
2\pi\left.\frac{d\sigma}{d\Delta\eta dR d\Delta\Phi}
\right/\frac{d\sigma}{d\Delta\eta dR}=
1+ \nonumber 
\frac{2}{\sigma_0(\Delta\eta,R)}\sum_{p=1}^\infty \sigma_p(\Delta\eta,R) \cos(p\Delta\Phi)
\nonumber 
\end{eqnarray}
where in the NLL BFKL framework,
\begin{eqnarray}
\sigma_p&=& \int_{E_T}^\infty \frac{dQ}{Q^3}
\alpha_s(Q^2/R)\alpha_s(Q^2R) \nonumber 
\left( \int_{y_<}^{y_>} dy x_1 f_{eff}(x_1,Q^2/R)x_2f_{eff}(x_2,Q^2R) 
\right) \nonumber
\\
&~& \int_{1/2-\infty}^{1/2+\infty}\frac{d\gamma}{2i\pi}R^{-2\gamma}
\ e^{\bar\alpha(Q^2)\chi_{eff}(p, \gamma, \bar{\alpha})\Delta\eta}  \nonumber
\end{eqnarray}
and $\chi_{eff}$ is the effective resummed kernel.
Computing the different $\sigma_p$ at NLL for the resummation schemes S3 and S4
allowed us to compute the azimuthal correlations at NLL. As expected, the
$\Delta \Phi$ dependence is less flat than for BFKL LL and is closer to the
DGLAP behaviour \cite{mnjet}. In Fig.~\ref{Figlhc}, we display the observable $1/\sigma d \sigma/d\Delta \Phi$ as a function
of $\Delta\Phi$, for LHC kinematics. The results are displayed for different values of 
$\Delta\eta$ and at both LL and NLL accuracy using the S4 resummation scheme. In general, the 
$\Delta\Phi$ spectra are peaked around $\Delta\Phi\!=\!0,$ which is indicative of jet emissions occuring back-to-back. 
In addition the $\Delta\Phi$ distribution flattens with increasing 
$\Delta\eta\!=\!y_1\!-\!y_2$. Note the change of scale on the vertical axis 
which indicates the magnitude of the NLL corrections with respect to the LL-BFKL results. The NLL corrections 
slow down the azimuthal angle decorrelations for both increasing $\Delta\eta$ and $R$ deviating from $1.$ We also
studied the $R$ dependence of our prediction which is quite weak~\cite{mnjet}
and the scale dependence
of our results by modifying the scale $Q^2$ to either $Q^2/2$ or $2Q^2$ and the effect on the azimuthal distribution 
is of the order of 20\%. The effect of the energy conservation in the BFKL equation~\cite{mnjet} is large when $R$ goes
away from 1. The effect is to reduce the effective value of $\Delta \eta$ between the jets and thus the decorrelation 
effect. However, it is worth noticing that this effect is negligible when $R$ is close to 1 where this measurement 
will be performed.

A measurement of the cross-section 
$d\sigma^{hh\!\to\!JXJ}/d\Delta\eta dR d\Delta\Phi$ at the Tevatron (Run 2) or the LHC will allow for a 
detailed study of the BFKL QCD dynamics since the DGLAP evolution leads to much less jet angular
decorrelation (jets are back-to-back when $R$ is close to 1). In particular, measurements with 
values of $\Delta\eta$ reaching 8 or 10 will be of great interest, as these could allow to distinguish between 
BFKL and DGLAP resummation effects and would provide important tests for the relevance of the BFKL formalism. 

To illustrate this result, we give in Fig.~\ref{Fig3} the azimuthal
correlation in the CDF acceptance. The CDF collaboration installed the
mini-Plugs calorimeters aiming for rapidity gap selections in the very forward
regions and these detectors can be used to tag very forward jets. A measurement
of jet $p_T$ with these detectors would not be possible but their azimuthal
segmentation allows a $\phi$ measurement. In Fig.~\ref{Fig3}, we display the jet
azimuthal correlations for jets with a $p_T>5$ GeV and $\Delta \eta=$6, 8, 10
and 11. For $\Delta \eta=$11, we notice that the distribution is quite flat,
which would be a clear test of the BFKL prediction.

\section{Jet gap jets at the Tevatron and the LHC}

In this section, we describe another possible measurement which can probe BFKL
resummation effects and we compare our predictions with existing D0 and CDF
measurements~\cite{usb}.

\subsection{BFKL NLL formalism}

The production cross section of two jets with a gap in rapidity between them reads
\begin{equation}
\frac{d \sigma^{pp\to XJJY}}{dx_1 dx_2 dE_T^2} = {\cal S}f_{eff}(x_1,E_T^2)f_{eff}(x_2,E_T^2)
\frac{d \sigma^{gg\rightarrow gg}}{dE_T^2},
\label{jgj}\end{equation}
where $\sqrt{s}$ is the total energy of the collision,
$E_T$ the transverse momentum of the two jets, $x_1$ and $x_2$ their longitudinal
fraction of momentum with respect to the incident hadrons, $S$ the survival probability,
and $f$ the effective parton density functions~\cite{usb}. The rapidity gap
between the two jets is $\Delta\eta\!=\!\ln(x_1x_2s/p_T^2).$ 

The cross section is given by
\begin{equation}
\frac{d \sigma^{gg\rightarrow gg}}{dE_T^2}=\frac{1}{16\pi}\left|A(\Delta\eta,E_T^2)\right|^2
\end{equation}
in terms of the $gg\to gg$ scattering amplitude $A(\Delta\eta,p_T^2).$ 

In the following, we consider the high energy limit in which the rapidity gap $\Delta\eta$ is assumed to be very large.
The BFKL framework allows to compute the $gg\to gg$ amplitude in this regime, and the result is 
known up to NLL accuracy
\begin{equation}
A(\Delta\eta,E_T^2)=\frac{16N_c\pi\alpha_s^2}{C_FE_T^2}\sum_{p=-\infty}^\infty\intc{\g}
\frac{[p^2-(\g-1/2)^2]\exp\left\{\bar\alpha(E_T^2)\chi_{eff}[2p,\g,\bar\alpha(E_T^2)] \Delta \eta\right\}}
{[(\g-1/2)^2-(p-1/2)^2][(\g-1/2)^2-(p+1/2)^2]} 
\label{jgjnll}\end{equation}
with the complex integral running along the imaginary axis from $1/2\!-\!i\infty$ 
to $1/2\!+\!i\infty,$ and with only even conformal spins contributing to the sum, and 
$\bar{\alpha}=\alpha_S N_C/\pi$ the running coupling.

Let us give some more details on formula \ref{jgjnll}. The NLL-BFKL effects are 
phenomenologically taken into account by the effective kernels $\chi_{eff}(p,\g,\bar\alpha)$.
The NLL 
kernels obey a {\it consistency condition} which allows to reformulate the 
problem in terms of $\chi_{eff}(\g,\bar\alpha).$ The effective kernel
$\chi_{eff}(\g,\bar\alpha)$ is obtained from the NLL kernel $\chi_{NLL}\lr{\g,\omega}$ by 
solving the implicit equation
$\chi_{eff}=\chi_{NLL}\lr{\g,\bar\alpha\ \chi_{eff}}$ as a solution of the 
consistency condition as it was also performed for forward jets.

In this study, we performed a parametrised distribution of $d \sigma^{gg\rightarrow gg}/dE_T^2$
so that it can be easily implemented in the Herwig Monte Carlo~\cite{herwig} since performing the integral over
$\gamma$ in particular would be too much time consuming in a Monte Carlo. The implementation of the
BFKL cross section in a Monte Carlo is absolutely necessary to make a direct comparison with data.
Namely, the measurements are sensititive to the jet size (for instance, experimentally the gap size
is different from the rapidity interval between the jets which is not the case by definition in the
analytic calculation).

\begin{figure}
\begin{center}
\epsfig{file=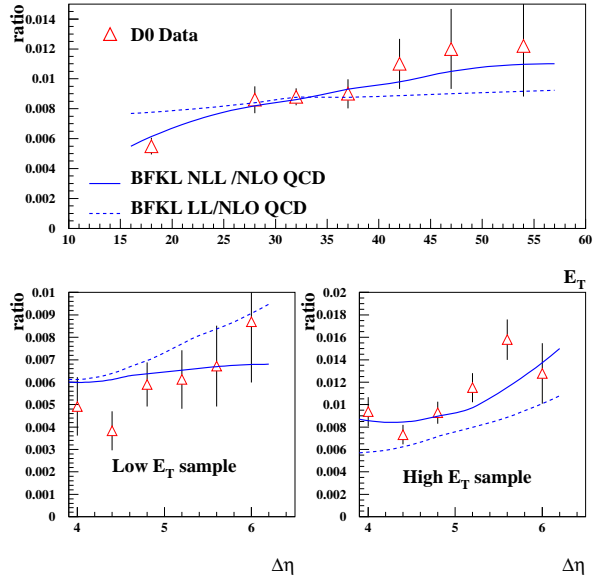,width=8.5cm}
\caption{Comparisons between the D0 measurements of the jet-gap-jet event ratio with the NLL- 
and LL-BFKL calculations. The NLL calculation is in fair agreement with the data. 
The LL calculation leads to a worse description of the data.}
\label{d0}
\end{center}
\end{figure}

\begin{figure}
\begin{center}
\epsfig{file=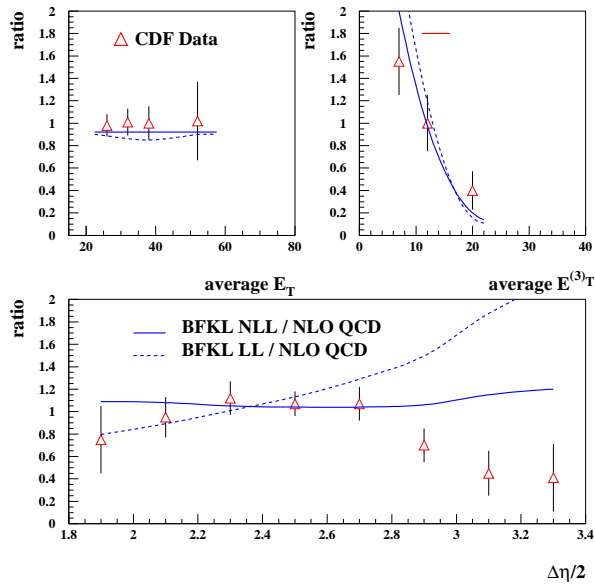,width=8.5cm}
\caption{Comparisons between the CDF measurements of the jet-gap-jet event ratio with the NLL- 
and LL-BFKL calculations. The NLL calculation is in fair agreement with the data. 
The LL calculation leads to a worse description of the data.}
\label{cdf}
\end{center}
\end{figure}

\subsection{Comparison with D0 and CDF measurements}
Let us first notice that the sum over all conformal spins is absolutely necessary. Considering
only $p=0$ in the sum of Equation~\ref{jgjnll} leads to a wrong normalisation and a wrong jet $E_T$
dependence, and the effect is more pronounced as $\Delta \eta$ diminishes.

The D0 collaboration measured the jet gap jet cross section ratio with respect to the total dijet
cross section, requesting for a gap between -1 and 1 in rapidity, as a function of the second
leading jet $E_T$,
and $\Delta \eta$ between the two leading jets for two different low and high $E_T$ samples
(15$<E_T<$20 GeV and $E_T>$30 GeV). To compare with theory, we compute the following quantity
\begin{eqnarray}
Ratio = \frac{BFKL~ NLL~HERWIG}{Dijet~Herwig} \times \frac{LO~QCD}{NLO~QCD} 
\end{eqnarray}
in order to take into account the NLO corrections on the dijet cross
sections, where $BFKL~ NLL$ $HERWIG$ and $Dijet~Herwig$ denote the BFKL NLL and the dijet cross section
implemented in HERWIG. The NLO QCD cross section was computed using the NLOJet++ program~\cite{nlojet}.

The comparison with D0 data~\cite{d0jgj} is shown in Fig. 5. We find a good agreement between the data
and the BFKL calculation. It is worth noticing that the BFKL NLL calculation leads to a better result
than the BFKL LL one (note that the best description of data is given by the BFKL LL formalism 
for $p=0$ but it does not make sense theoretically to neglect the higher spin components and this
comparison is only made to compare with previous LL BFKL calculations).

The comparison with the CDF data~\cite{d0jgj} as a function of the average jet $E_T$ and the
difference in rapidity between the two jets is shown in Fig. 6, and the conclusion remains the same:
the BFKL NLL formalism leads to a better description than the BFKL LL one.

\subsection{Predictions for the LHC}
Using the same formalism, and assuming a survival probability of 0.03 at the LHC, it is possible to
predict the jet gap jet cross section at the LHC. While both LL and NLL BFKL formalisms lead to a
weak jet $E_T$ or $\Delta \eta$ dependence, the normalisation is found to be
quite different (see Fig. 7)
leading to higher cross section for the BFKL NLL formalism.

\begin{figure}
\begin{center}
\epsfig{file=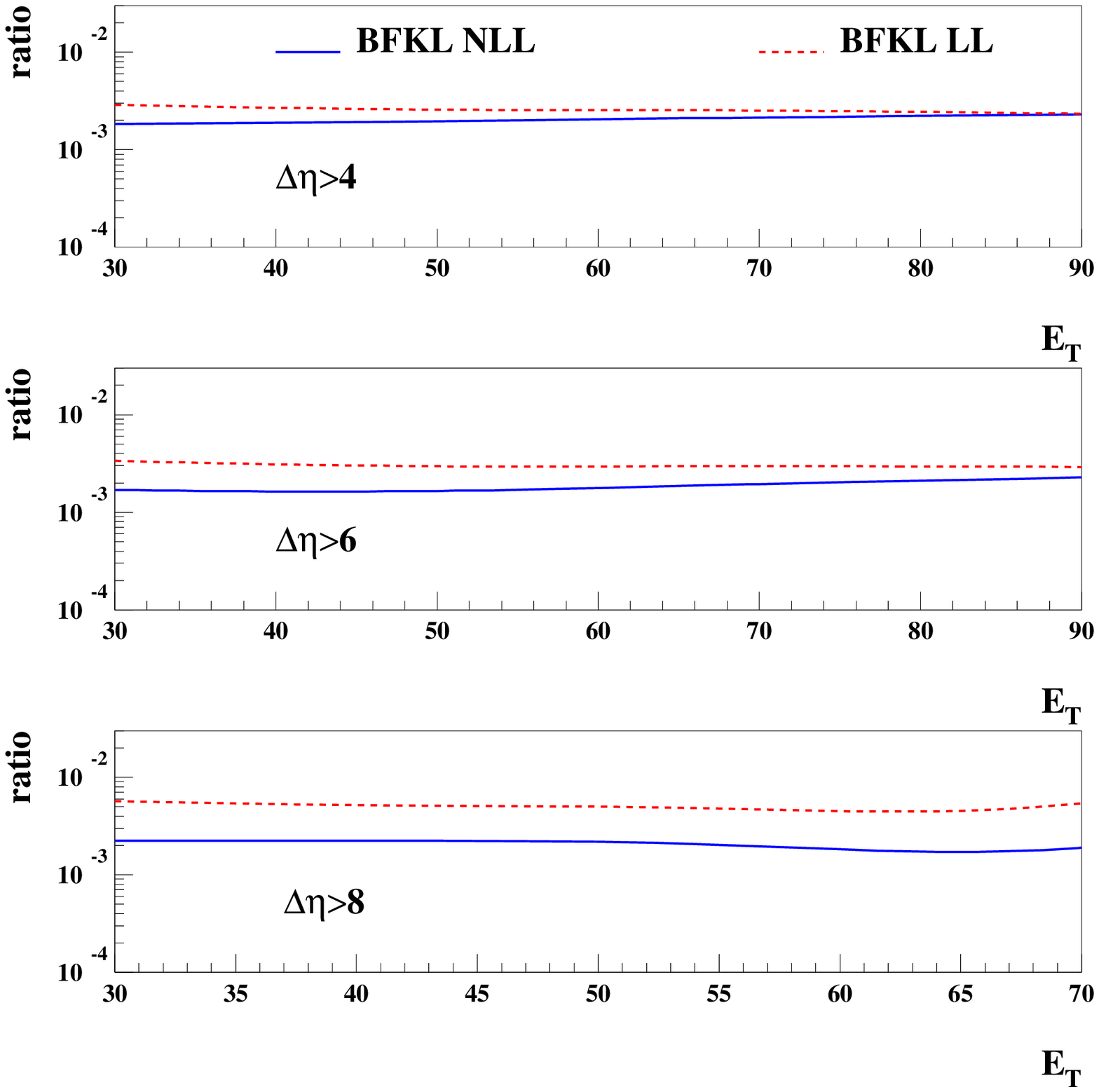,width=8.5cm}
\hfill
\epsfig{file=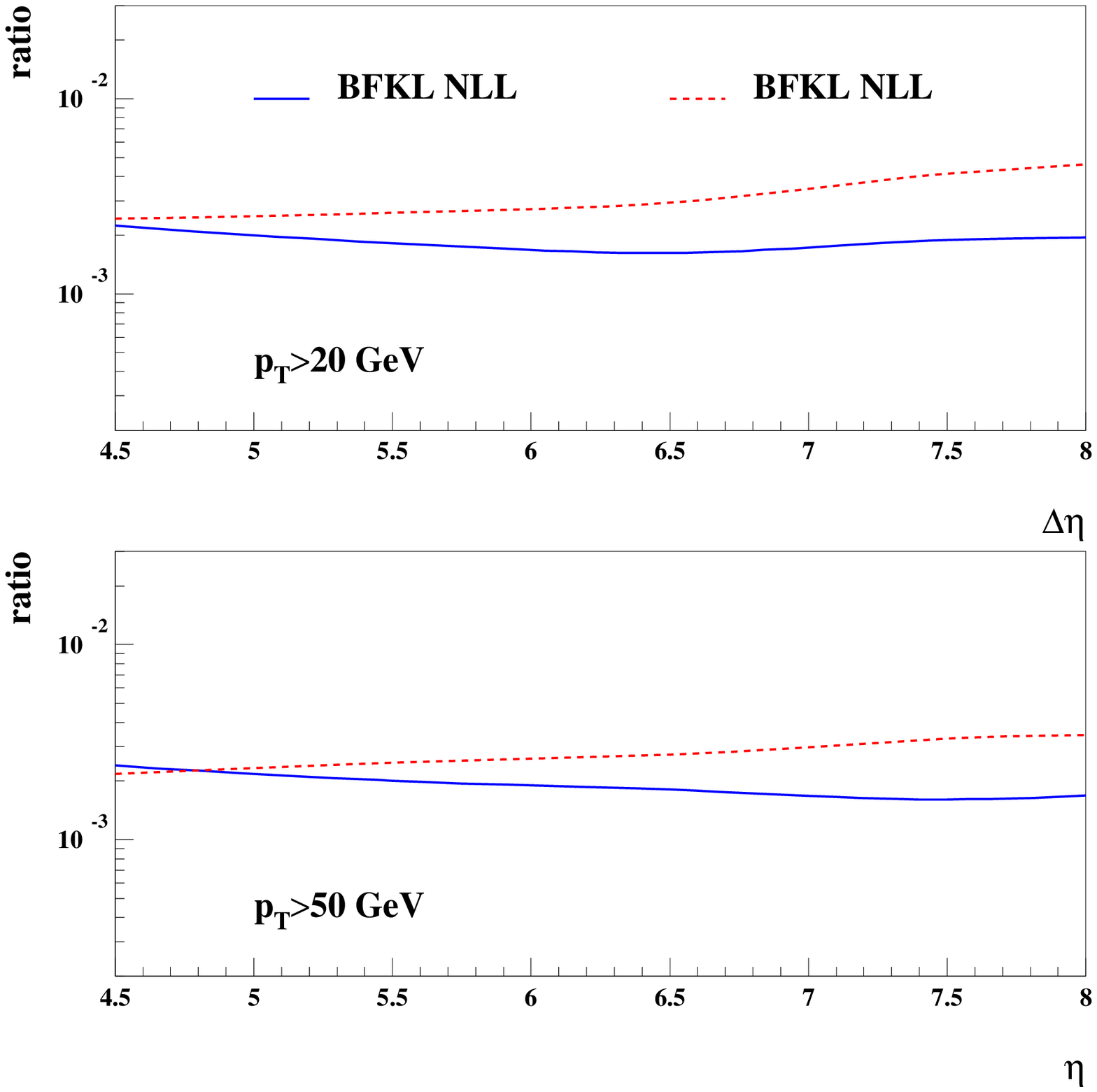,width=8.5cm}
\caption{Ratio of the jet gap jet to the inclusive jet cross sections at the LHC as a function of jet $p_T$ and $\Delta \eta$.}
\label{fig6}
\end{center}
\end{figure}

\section{Quartic anomalous couplings at the LHC}

In the third part of this report, we discuss a completely different topic,
namely the possibility to probe anomalous quartic couplings between photons and
$W$ or $Z$ bosons at the LHC with an unprecedent precision using forward
detectors to be installed in CMS and ATLAS experiments~\cite{us}.
In the Standard Model (SM) of particle physics, the couplings of fermions and 
gauge bosons are constrained by the gauge symmetries of the Lagrangian.
The measurement of $W$ and $Z$ boson pair productions via the exchange of
two photons allows to provide directly stringent tests
of one of the most important and least understood
mechanism in particle physics, namely the
electroweak symmetry breaking~\cite{stirling}. The non-abelian gauge nature of the SM
predicts the existence of quartic couplings
$WW\gamma \gamma$
between the $W$ bosons and the photons which can be probed directly at the 
Large Hadron Collider (LHC) at CERN.
The quartic coupling to the $Z$ boson $ZZ\gamma \gamma$ is not present in the
SM. Quartic anomalous couplings between the photon and the $Z$ or $W$ bosons
are specially expected to occur in higgsless or extradimension
models~\cite{higgsless}.

%%%%%%%%%%%%%%%%%%%%%%%%%%%%%%%%%%%%%%%%%%%%%%%%%%%%%%%%%%%%%%%%%
\subsection{Photon exchange processes in the SM}
%%%%%%%%%%%%%%%%%%%%%%%%%%%%%%%%%%%%%%%%%%%%%%%%%%%%%%%%%%%%%%%%%%
The process that we intend to study is the $W$ pair production shown in Fig.~8
induced by the 
exchange of two photons~\cite{piotr,us}. It is a pure QED process
in which the decay products of the $W$ bosons are measured in the central 
detector and the scattered protons leave intact in
the beam pipe at very small angles, contrary to inelastic collisions. Since 
there is no proton remnant the process is purely exclusive; only $W$ decay products 
populate the central detector, and the intact protons can be detected in
dedicated detectors located along the beam line far away from the interaction
point.

The cross section of the $pp\rightarrow p WW p$ process which proceeds through 
two-photon exchange is calculated as a convolution of the 
two-photon luminosity and the total cross section $\gamma\gamma\rightarrow WW$.
The total two-photon cross section is 95.6 fb. 

All considered processes (signal and background) were produced using the Forward
Physics Monte Carlo~\cite{fpmc} (FPMC) generator. The aim of FPMC is to produce different
kinds of processes such as inclusive and exclusive diffraction, photon-exchange
processes. FPMC was interfaced to as fast simulation of the ATLAS
detector~\cite{atlfast}. To reduce the amount of considered background, we only
use leptonic (electrons and muons) decays of $Z$ and $W$ bosons.  The following
backgrounds were considered:
%\begin{enumerate}
$\gamma\gamma\rightarrow l\bar{l}$ --- two-photon dilepton production,
DPE$\rightarrow l\bar{l}$ ---- dilepton production through double pomeron 
exchange,
DPE$\rightarrow W^+W^-\rightarrow l\bar{l}\nu\bar{\nu}$ --- diboson 
production through double pomeron exchange.
%\end{enumerate}

After simple cuts to select exclusive $W$ pairs decaying into leptons, such
as a cut on the proton momentum loss of the proton ($0.0015<\xi<0.15$) --- we
assume the protons to be tagged in the ATLAS Forward Physics
detectors~\cite{afp} ---,
on the transverse momentum of the leading and second leading leptons at 25 and
10 GeV respectively, on $\met>20$ GeV, $\Delta \phi>2.7$ between leading
leptons, and $160<W<500$ GeV, the diffractive mass reconstructed using the
forward detectors, the background is found to be less than 1.7 event for 30
fb$^{-1}$ for a SM signal of 51 events. In this channel, a 5 $\sigma$ discovery
of the Standard Model $pp\rightarrow pWWp$ process is possible after 5 fb$^{-1}$.  

\begin{figure}
\begin{center}
\epsfig{file=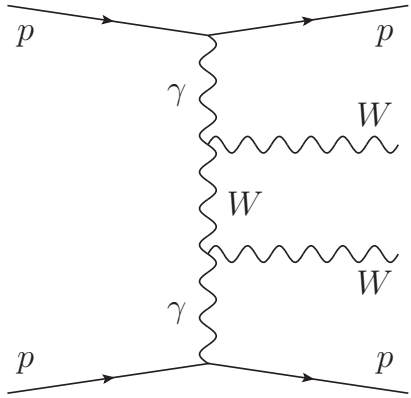,width=6.cm} 
\end{center}
\caption{Sketch diagram showing the two-photon production of a central system.}
\end{figure}

\begin{figure}
\begin{center}
\epsfig{file=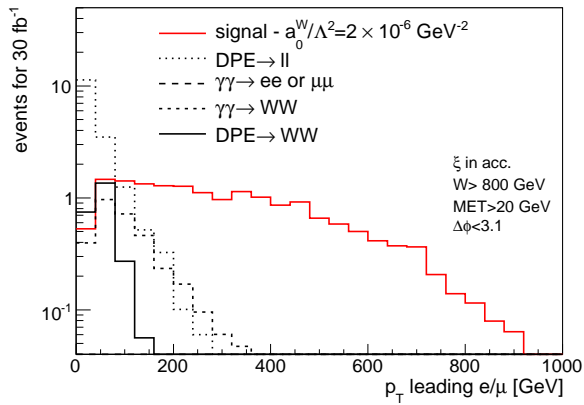,width=8.cm} 
\end{center}
\caption{Distribution of the transverse momentum of the leading lepton for
signal and background after the cut on $W$, $\met$, and $\Delta \phi$ between
the two leptons.}
\end{figure}

\subsection{Quartic anomalous couplings}
The parameterization of the quartic couplings
based on \cite{Belanger:1992qh} is adopted. We concentrate on the lowest order 
dimension operators which have
the correct Lorentz invariant structure and obey the $SU(2)_C$ 
custodial symmetry in order to fulfill the stringent experimental bound on the 
$\rho$ parameter. 
The lowest order interaction
Lagrangians which involve two photons are dim-6 operators. 
The following expression for the effective quartic Lagrangian is used
 \begin{eqnarray}
     \mathcal{L}_6^0 &=& \frac{-e^2}{8} \frac{\aZerow}{\Lambda^2} F_{\mu\nu} 
     F^{\mu\nu} W^{+\alpha} W^-_\alpha 
     - \frac{e^2}{16\cos^2 \theta_W} \frac{a^Z_0}{\Lambda^2} F_{\mu\nu} 
     F^{\mu\nu} Z^\alpha
Z_\alpha\nonumber \\
     \mathcal{L}_6^C & = & \frac{-e^2}{16} \frac{\aCw}{\Lambda^2} 
     F_{\mu\alpha} F^{\mu\beta} (W^{+\alpha} W^-_\beta + W^{-\alpha} 
     W^+_\beta) 
	- \frac{e^2}{16\cos^2 \theta_W} \frac{a^Z_C}{\Lambda^2} 
	F_{\mu\alpha} F^{\mu\beta} Z^\alpha Z_\beta
\label{eq:anom:lagrqgc}
\end{eqnarray}
where $a_0$, $a_C$ are the parametrized new coupling constants and the new scale $\Lambda$ is introduced
so that the Lagrangian density has the correct dimension four and is 
interpreted as the typical mass scale of  new
physics.
In the above formula, we allowed the $W$ and $Z$ parts of the Lagrangian to 
have specific couplings, i.e.
 $a_0\rightarrow (\aZerow$, $a^Z_0$) and similarly $a_C\rightarrow(\aCw$, 
 $\aCz$).

The $WW$ and $ZZ$ two-photon cross sections rise quickly at high energies when 
any of the anomalous parameters are non-zero. The cross section rise has to be 
regulated by a form factor which vanishes in the high energy limit to 
construct a realistic physical model of the BSM theory. We therefore 
modify the couplings by form factors 
that have the desired behavior, i.e. they modify the coupling at small 
energies only slightly but suppress it  when the center-of-mass energy $W_{\gamma\gamma}$ 
increases. The form of the form factor that we consider is the following
\begin{eqnarray}
a\rightarrow \frac{a}{(1+W^2_{\gamma\gamma}/\Lambda^2)^n}
\label{eq:anom:formfactor}
\end{eqnarray}
where $n$=2, and $\Lambda \sim$2 TeV.

The cuts to select quartic anomalous gauge coupling $WW$ events are similar as the
ones we mentioned in the previous section, namely $0.0015<\xi<0.15$ for the
tagged protons, $\met>$ 20 GeV, $\Delta \phi<3.13$ between the two leptons. In
addition, a cut on the $p_T$ of the leading lepton $p_T>160$ GeV and on the
diffractive mass $W>800$ GeV are requested since anomalous coupling events
appear at high mass. Fig~9 displays the $p_T$ distribution of the leading lepton
for signal and the different considered backgrounds. 
After these requirements, we expect about 0.7 background
events for an expected signal of 17 events if the anomalous coupling is about
four order of magnitude lower than the present LEP limit ($|a_0^W / \Lambda^2| =
5.4$ 10$^{-6}$) for a luminosity of 30 fb$^{-1}$. The strategy to select anomalous coupling $ZZ$ events is
analogous and the presence of three leptons or two like sign leptons are 
requested. 
Table 1 gives the reach on anomalous couplings at the LHC for a
luminosity of 30 and 200 fb$^{-1}$ compared to the present OPAL limits~\cite{opal}. We note that we can gain almost
four orders of magnitude in the sensitivity to anomalous quartic gauge couplings
compared to LEP experiments, and it is possible to reach the values expected in Higgsless
or extra-dimension models which are of the order of 5 10$^{-6}$. 
The tagging of the protons using the ATLAS Forward
Physics detectors is the only method at present to test such small values of
quartic anomalous couplings and thus to probe the higgsless models in a clean
way. The reach on anomalous triple gauge couplings is much less improved at the
LHC compared to LEP experiments~\cite{kepka}.

To conclude, the ATLAS Forward Physics program (and the CMS one) will allow to
study Higgsless models with an unprecedent precision as well as to probe the
Higgs boson by allowing its mass and spin measurements~\cite{higgs} using the
forward detectors proposed for installation at 220 and 420 m in ATLAS and CMS.

\begin{table}
\begin{center}
   \begin{tabular}{|c||c|c|c|}
    \hline
    %\raisebox{-1.5ex}[0pt][0pt] 
    Couplings & 
    OPAL limits & 
    \multicolumn{2}{c|}{Sensitivity @ $\mathcal{L} = 30$ (200) fb$^{-1}$} \\
    &  \small[GeV$^{-2}$] & 5$\sigma$ & 95\% CL \\ 
    \hline
    $a_0^W/\Lambda^2$ & [-0.020, 0.020] & 5.4 10$^{-6}$ & 2.6 10$^{-6}$\\
                      &                 & (2.7 10$^{-6}$) & (1.4 10$^{-6}$)\\ \hline               
    $a_C^W/\Lambda^2$ & [-0.052, 0.037] & 2.0 10$^{-5}$ & 9.4 10$^{-6}$\\
                      &                 & (9.6 10$^{-6}$) & (5.2 10$^{-6}$)\\ \hline               
    $a_0^Z/\Lambda^2$ & [-0.007, 0.023] & 1.4 10$^{-5}$ & 6.4 10$^{-6}$\\
                      &                 & (5.5 10$^{-6}$) & (2.5 10$^{-6}$)\\ \hline               
    $a_C^Z/\Lambda^2$ & [-0.029, 0.029] & 5.2 10$^{-5}$ & 2.4 10$^{-5}$\\
                      &                 & (2.0 10$^{-5}$) & (9.2 10$^{-6}$)\\ \hline               
    \hline
   \end{tabular}
\end{center}
\caption{Reach on anomalous couplings obtained in $\gamma$ induced processes
after tagging the protons in the final state in the ATLAS Forward Physics
detectors compared to the present OPAL limits. The $5\sigma$ discovery and 95\%
C.L. limits are given for a luminosity of 30 and 200 fb$^{-1}$} 
\end{table}

\end{document}